\documentclass[3p,10pt,twocolumn]{elsarticle}

\usepackage{tcolorbox}
\usepackage{color,soul}
\usepackage{amssymb}
\biboptions{sort&compress} 
\usepackage{amsmath}
\usepackage{graphicx}
\usepackage{amsfonts}%
\usepackage{booktabs} % Allows the use of \toprule, \midrule and \bottomrule in tables
\usepackage{multirow}% always when we are using multirow(very important to center the 
\usepackage{xcolor}

\begin{document}

\begin{frontmatter}
\title{First principles prediction of the  solar cell efficiency of  chalcopyrite materials AgMX$_2$ (M=In,Al; X=S, Se,Te)}
\author[1]{GM Dongho-Nguimdo\corref{mycorrespondingauthor}} \cortext[mycorrespondingauthor]{Corresponding author} \ead{donghomoise@gmail.com} \author[1]{Emanuel Igumbor}  \author[2]{Serges Zambou} \author[3]{ Daniel P. Joubert }

\address[1]{School of Interdisciplinary Research and Graduate Studies, University of South Africa, 
 Pretoria, South Africa}   
  \address[2]{Tampere University, Division of Electronics and Telecommunications, Korkeakoulunkatu 3, FI-33720 Tampere, Finland}
\address[3]{The National Institute for Theoretical Physics, School of Physics and Mandelstam Institute for Theoretical
Physics, University of the Witwatersrand, Johannesburg, Wits 2050, South Africa} 
%\newpage
\begin{abstract}
Using the Spectroscopic Limited Maximum Efficiency,  and  Shockley and Queisser 
predictor models, we compute the solar efficiency of the chalcopyrites AgMX$_2$ (M=In,Al; 
X=S,Se,Te). The results presented are based on  the estimation of the electronic and optical 
properties obtained from first principles density functional theory as well as the  many-body 
perturbation theory calculations. The  results from this report were consistent with the 
experimental data.The optical bandgap was accurately  estimated  from the 
absorption spectra, obtained by solving the Bethe and Salpeter equation. Fitting the Tauc's plot 
on the absorption spectra, we also predicted that the materials studied have a  direct allowed 
optical transition. The theoretical estimations of the solar cell performance showed that the 
efficiencies from the Shockley and Queisser model are higher than those from the spectroscopic 
limited maximum efficiency model. This improvement is attributed to the absorption, the 
recombination processes and the optical transition accounted in the calculation of the efficiency.
%\textcolor{red}{expamd the abstract}
\end{abstract}

\end{frontmatter}

\section{Introduction}
Chalcopyrite materials have recently been investigated for their potential technological 
applications,
including the miniaturisation of electronic components and the harvesting of the solar energy 
\cite{xu2009,zunger1983,finland,alonso2001}.
They can be useful for application in non linear optics process including second harmonic 
generation~\cite{boyd1972linear}
as well as being used as thermoelectric materials~\cite{thermo}.
Considerable efforts have been 
made by researchers during the last decade to improve the performance of  solar cells. In 2018, 
Contreras et\textit{al.}~\cite{contreras}  were able to achieve a  19$\%$ efficiency using a tandem 
of ZnO/CdS/CuInSe$_2$. 
 Chalcopyrites are predicted to have a life-time in outer space fifty times longer than that of  
 silicon or III-V semiconductors~\cite{solar}.  Furthermore,  the researchers obtained  a very 
high stability against  electron   and  proton  irradiation in spatial applications. It is therefore 
important to screen other chalcopyrite materials other than the Copper-Indium-Sulphur family  to 
ascertained the existence of high efficient solar cell materials. Nowadays, due to high 
improvement in code development,  computational tools to perform such a study have 
reached the level of accuracy that allows an optimal and reliable screening. In this work, we focus 
 on two families of chalcopyrites, namely the AgMX$_2$ (M=In, Al; X=S, Se,Te). The direct 
bandgap of the AgMX$_2$  could be an advantage in the solar cell manufacturing. In addition, wide 
% \hl{bandgap such as AgAlX$_2$  could be useful in  making  multi-junction cells.  while the 
% AgInX$_2$  are preferable for the single junction.} 
Reports have shown that properties such as   the 
structural stability, size of the  and nature (whether it is direct or indirect) 
of the bandgap, dielectric function, energy loss and absorption coefficient of a material are  the 
important parameters which are used to predict a given material as  good solar cell absorber. 
Theoretical efficiency calculation is a step ahead of the first principles methods in the search for 
potential solar cell materials. Numerous predictor models including the Shockley and Queisser 
(SQ)~\cite{sq},  Spectroscopic Limited Maximum Efficiency (SLME)~\cite{liping} and Spectroscopic 
limited practical efficiency (SLPE)~\cite{slpe} have been used to tackle this problem. 
In this work, we use the SQ and SLME models to show that the efficiency  of the studied materials does not only depend on the bandgap, but also the physics of the absorption and the recombination processes as well as the nature of the dipole transition. 
The outline of this paper is as follows: in Section 2, computational details are presented, Section 3  is dedicated to the electronic and the optical properties of the materials, in Section 4 we present the result of the theoretical efficiency calculations of the materials and the work is summarized in Section 5.

\section{Computational details}
\subsection{first principles calculations}
The results of this work  were performed by means of  density functional theory as 
implemented in the Vienna 
\textit{ab-initio}  Simulation Package (VASP)~\cite{VASP}.  Electron-ion interaction was mimiced by 
the projector-augmented wave formalism~\cite{KresseJoubert}. Since our main interest is not on the 
structural parameters,
we used already optimized structural parameters from previous studies~\cite{paper1,paper2}
where the generalised gradient approximation (GGA) in the revised Perdew-Burke-Ernzerhof version 
for solids (PBEsol)~\cite{PBEsol} was used as exchange-correlation functional.
 Convergence tests showed that the Monkhorst-Pack~\cite{monkhorst} \textbf{k}-points 
 mesh of $7\times 7\times7$ with an energy cut-off of 550 eV were sufficient  for sampling  the 
Brillouin zone. 
The geometric structures were  relaxed until the  final 
change in the
total energy was less than 10$^{-5}$ eV and the forces acting on the atoms were relaxed to
below 0.001 eV/$\mathrm{\AA}$. Accurate description of the electronic properties is important for
calculating the  optical properties.  We first used a single short GW~\cite{GW} also known as 
G$_0$W$_0$, where the quasi-particle energies are from one GW iteration follow by a semi 
self-consistent GW$_0$ where only the Green's function is updated. In the G$_0$W$_0$ calculations, 
the quasiparticles energies, $\epsilon^{QP}_{n\textbf{k}}$, are solution of the linear equation  
obtained by a Taylor expansion of the self-energy, $\Sigma$, around the DFT energies  and all the 
off-diagonal matrix elements are neglected~\cite{self-energy}:
 \begin{align}\label{eqr1}
 \epsilon^{QP}_{n\textbf{k}}=\epsilon_{n\textbf{k}}+Z_{n\textbf{k}}\left[Re\Sigma_{n\textbf{k}}(\epsilon_{n\textbf{k}})-V_{n\textbf{k}}^{xc}\right],
 \end{align}
 where $\epsilon_{n\textbf{k}}, \Sigma_{n\textbf{k}}$, $V_{n\textbf{k}}^{xc}$ and $Z_{n\textbf{k}}$ 
are the Kohn-Sham
energies, the diagonal matrix element of $\Sigma$, the exchange-correlation potential and the 
renormalization factor, respectively. In practice, the Kohn-Sham eigenstates and eigenenergies in 
the calculation of  $\epsilon^{QP}_{n\textbf{k}}$  according to Equation\eqref{eqr1}  are usually  
taken from either the Hartree-Fock, a GGA, the local density approximation (LDA)  or an hybrid 
functional initial calculation. In order to increase the accuracy of our calculations, we used the 
hybrid functional (HSE06) exchange-correlation to obtain $\epsilon_{n\textbf{k}}$. This method helps 
to capture some errors such as wrong hybridization of orbitals, localization and delocalization 
errors encounter in GGA and LDA. 
The absorption coefficient is calculated by using the many-body perturbation theory at the Bethe and 
Selpeter level~\cite{bse1,bse2} built on top of the GW$_0$. 
 Hence, the two particles interaction kernel is constructed  and the BSE equation is solved in the 
Tamn-Dancoff 
approximation~\cite{abdulsalamTCX2}. After a set of convergence test, we set the following: the 
number of additional empty bands to 1008, the energy cutoff for the response function to 300 eV and 
the number of frequency grid points to 192 for the semi self-consistent calculations. Ten  occupied 
and unoccupied orbitals were  included in the BSE calculation in order to get accurate positions of 
the absorption peaks.

 \subsection{The SQ and the SLME models}
Using the first principles calculation method, the solar cell performances  were calculated 
through the SQ and SLME models. 
The efficiency  $\eta$, of a solar cell is the ratio of the maximum power delivered by the cell 
(P$_{max}$) and the incident solar power striking on the cell (P$_{in}$): $\eta=P_{max}/P_{in}.$
In the SQ model, it is assumed that each photon with energy above the bandgap of the absorber  
produces 
an electron-hole pair. Hence, the maximum output  per unit area per unit  time can be expressed as 
\begin{equation}
 P_{max}=E_gN_{ph}
\end{equation}

where E$_g$ is the bandgap of the absorber and N$_{ph}$ is the number of  incident photon  per unit 
area per unit time with energy above E$_g$. At a first approximation, the  N$_{ph}$ can be  
calculated using the Planck equation~\cite{theoriefficiency}. However, for a more reliable 
estimation, a common standard  solar spectra model used as reference to allow comparison of solar 
cell models and devices  is the air mass 1.5 global 
spectrum referred to as AM1.5G spectrum \cite{am15,am15g}. The AM1.5G accounts for the relative path 
length taken by the sun's rays through the atmosphere before reaching the 
ground~\cite{am15theory1,am15theory2}.
Considering that all electron are not always automatically converted into current due to the 
recombination process,
the net  current  is given by~\cite{slme2,phdthesis}:
 \begin{equation}
  J=J_{sc}-J'\left[ e^{(qV/k_BT_c)}-1\right],\label{j}
 \end{equation}

where $k_B$, $T_c$, $V$, and $q$ are the Boltzmann constant and the temperature of cell, the current 
voltage 
delivered by the module, and the electron charge, respectively. $J_{sc}$ is the short circuit 
current, and $J'$ is the current recombination rate. Contrary to the SQ model, the SLME model 
accounts for the photon absorptivity $a(E)$ whereby $J'$ is calculated by 
\begin{equation}
 J'=\beta\int_0^{\infty}\varepsilon(E)N_{ph}(E,T_c)dE,\label{jprime}
\end{equation}
   where  $\varepsilon(E)$  is the emittance and $\beta$ is a coefficient proportional to the 
fraction of the radiative electron-hole recombination  current. According to the ``principle of 
detailed balance''~\cite{sq}, absorptivity and emittance are equal and defined as 
$a(E)=1-e^{2\alpha(E)L}$ where L is the thickness of the absorber and $\alpha(E)$ the absorption 
coefficient from first principal calculations.

%\section{Results and discussions}
\section{Electronics and optical properties}
\begin{table}[hbtp]
\centering%
\caption{The results of the structural parameters adapted from Ref.~\cite{paper1,paper2}, 
comparison with experimental data is also provided.}\label{t1}
\resizebox{8.1cm}{!}{
\begin{tabular}{cccccccccc}\\
\toprule
Materials& Functional & a(\AA) & c/a & V$_0$(\AA$^3)$&B$_0$(GPa)    \\ \toprule
\multirow{2}*{AgAlTe$_2$} & PBEsol       &  6.31   &  1.906 & 29.91&49.83  \\
& Exp.      &  6.29    &  1.880 & 29.31 & -&\\	\midrule
\multirow{2}*{AgAlSe$_2$}  & PBEsol         &  5.91  &  1.866 & 24.00 & 61.23 \\
	 & Exp.      &  5.95    &  1.806 & 23.83 & -\\	\midrule		
\multirow{2}*{AgAlS$_2$}  &PBEsol    &  5.66   &  1.822 & 20.69 & 73.78 \\
& Exp.      &  5.72    &  1.770 & 20.86 & -\\

\midrule		
\multirow{2}*{AgInS$_2$}  &PBEsol    &  5.80 &  1.653 & 23.86 & 62.40 \\
& Exp.      &  5.81&  1.929 & 24.17 & 62\\
	\bottomrule			            	 
\end{tabular}
}
\end{table}

We calculated the size and the nature of the bandgap using  the optimized structural parameters as 
shown in Table~\ref{t1} as well as the experimental parameters.
In this report, Table~\ref{t2} lists the results as obtained from our theoretical calculations, 
other theoretical results as well as the experimental results.
The results show that the bandgaps obtained using the PBEsol are underestimated. This corroborate 
the results of  LDA and PBE functionals, which are known to underestimate the bandgap of 
materials. In contrast, the results from the modified Becke-Johnson (MBJ) which is a metaGGA 
functional are already in the range of the experimental results despite the relatively low 
computational time~\cite{mbj}. The hybrid functional HSE06 slightly overestimates the bandgap by an 
average of  $4.72\%$. However, the single short GW results are not close to the experimental data   
despite using the  HSE06 eigenvalues as initial input  and a large number of empty bands for those 
calculations. For example, the G$_0$W$_0$ predicts a bandgap of 2.12 and 1.5 eV for AgAlSe$_2$ and 
AgInS$_2$, respectively, while their experimental values are in that other   2.55 eV and 1.86 eV. 
The MBJ results are much improved than those from G$_0$W$_0$. 
calculations.
\begin{table}
[hbtp]
\centering%
\caption{Results of bandgaps using different functionals. MBJ (modified Becke-Johnson) is a metaGGA 
exchange-correlation functional. Comparison with experimental data is also provided. All the 
materials are predicted to have a direct bandgap.  $b$=Ref.~\cite{paper1} and 
$c$=Ref.~\cite{paper2}.}\label{t2}
\resizebox{8.1cm}{!}{
\begin{tabular}{cccccccccc}\\\toprule
%\toprule
Materials& G$_0$W$_0$ & GW$_0$&Exp$^{b,c}$. & HSE06$^b$ & MBJ$^c$ & PBEsol$^b$    \\ \toprule
\multirow{1}*{AgInS$_2$} &0.92 & 1.90&1.86 &1.92 &1.73  &0.27\\\midrule
\multirow{1}*{AgAlS$_2$} &2.67 & 3.21& 3.13 &3.34 &3.15  &1.83\\\midrule
\multirow{1}*{AgAlSe$_2$}& 2.12& 2.46& 2.55 &2.70 &2.38  &1.11\\	\midrule
\multirow{1}*{AgAlTe$_2$} &2.08 & 2.22& 2.27 &2.34&2.14  &1.03\\		\bottomrule			            	 
\end{tabular}
}
\end{table} 
G$_0$W$_0$ has a deficiency in predicting chalcopyrite bandgaps accurately. Similar discrepancies 
were found in Ref.~\cite{aguilera} for the band structure and optical properties of 
CuGaS$_2$. It was attributed to the fact that an important contribution to the Cu-\textit{d} 
orbitals at the upper most valence band leads to a strong hybridisation with the Ga-\textit{p} 
orbitals.  Furthermore, Similar results were reported by S. Botti in the case  of CuInS$_2$ and 
CuInSe$_2$ \cite{botti}  where the author obtained a bandgap  of 0.28 and 0.25 eV, respectively 
against  experimental bandgap of 1.54 eV and 1.05 eV. In this study, since the materials are from 
the same family of chalcopyrites, we argue that this underestimation of the bandgap by G$_0$W$_0$ 
probably originated from its inability to capture the \textit{p-d} hybridisation.  We then went a 
step forward by employing a semi self-consistent GW calculation (GW$_0$) where the screened Coulomb 
interaction W$_0$ remains at the  random phase approximation (RPA) level and the Green's function 
updated by using the quasiparticle energy from the single shot calculation. The results of the 
GW$_0$  are very close to the experimental results. For instance, a bandgap of 2.21 and 1.9 eV were 
predicted for AgAlTe$_2$ and AgInS$_2$  while their experimental values are 2.27 and 1.86 eV, 
respectively. It is important to point out that all calculations predict a direct bandgap for all 
the materials irrespective of the functional used.

\begin{figure}[hbtp]
\centering 
\includegraphics[scale=0.222]{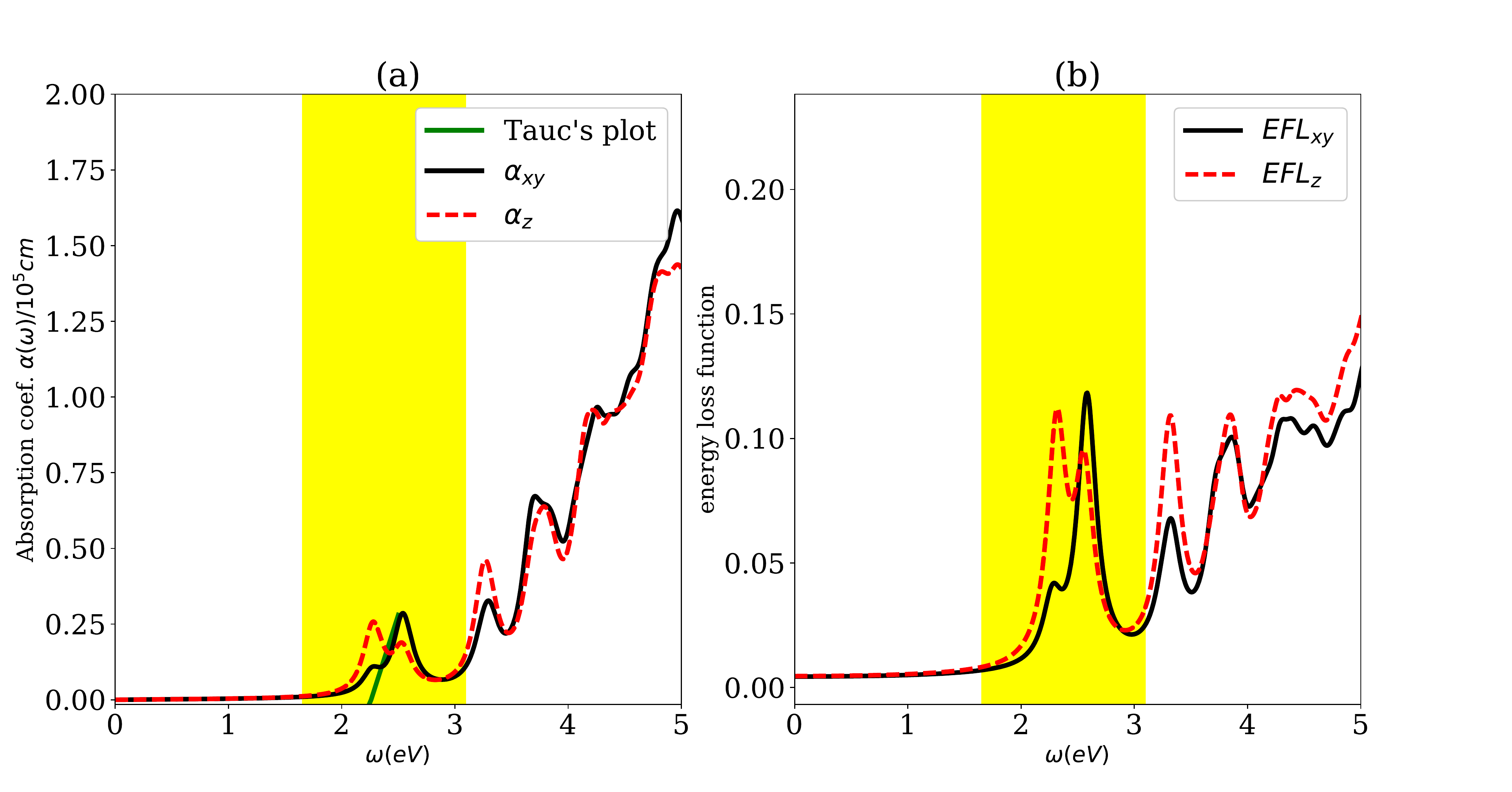}
\caption{Plots showing (a) average absorption spectra and (b) the energy loss function 
of AgAlSe$_2$ material. The shaded region represents the visible range of light (1.65-3.1 
eV)}\label{f1}
\end{figure}

\begin{figure}[hbtp]
\centering
\includegraphics[scale=0.4]{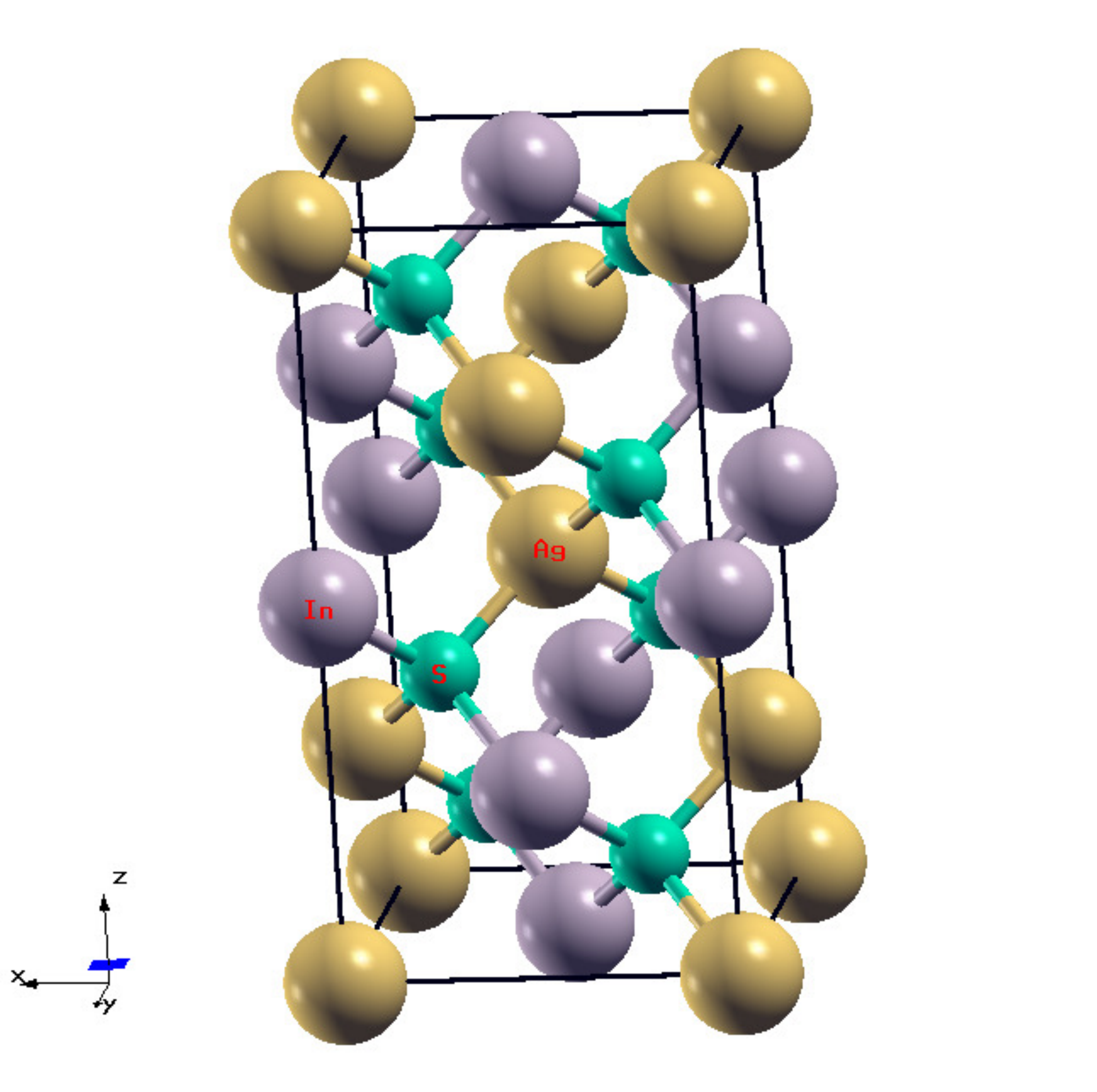}
\caption{Illustration of the chalcopyrites  structures $(a=b\approx c/2)$. The $xy$ plane is the perpendicular ($\perp$) while $z$  is the parallel ($\parallel$) direction.}\label{f11}
\end{figure}

Figure~\ref{f1}-a displays the  absorption spectra $\alpha(\omega)$ and energy loss
function from BSE calculations built on top of the semi self-consistent GW$_0$. The shaded region 
indicated the visible range where photon are absorbed.  
Because of the anisotropy nature of the materials as depicted in Figure~\ref{f11},
we plotted the absorption and the energy loss function (ELF) perpendicular to $xy$ plane and 
parallel to the $z$ direction.
These spectra are  similar to those from chalcopyrites of CuInS$_2$ and CuGaS$_2$ from  
Ref.~\cite{finland,alonso2001,soni2010} where theoretical and experimental techniques where 
used. This trend is  also  similar for other materials in this present report. For 
instance, the onset of absorption in the visible range is at 2.05 and 1.92 eV along the 
perpendicular($\perp$)  and parallel($\parallel$) direction, respectively. Two sharp absorption 
peaks appear in the visible range in each direction. The maximum  occurs  2.56  and 2.28 eV for an 
absorption  of $0.28\times10^5~cm^{-1}$ and $0.25\times10^5~cm^{-1}$   along the $\perp$ and 
$\parallel$ direction, respectively. The AgInS$_2$  has the highest absorption in the visible range 
suggesting it could make a better solar materials than the others.
It is worth  pointing out that for all the compounds under investigation,
these peaks occur at an energy lower that the fundamental bandgap. Hence, attesting the fact that 
our BSE calculations were able to capture the electron-hole interaction unlike the GW and the 
traditional DFT functionals.
We also calculated the EFL of the structures.  From Figure~\ref{f1}-b, 
we can observed that the ELF has plasmom peaks at the visible range. The position of the plasmon 
peaks shift toward lower energy in the structure as one moves down the chalcogenide group in the 
periodic table.

\section{Solar cell performance}
\begin{figure}[hbtp]	
\centering
\includegraphics[scale=0.42]{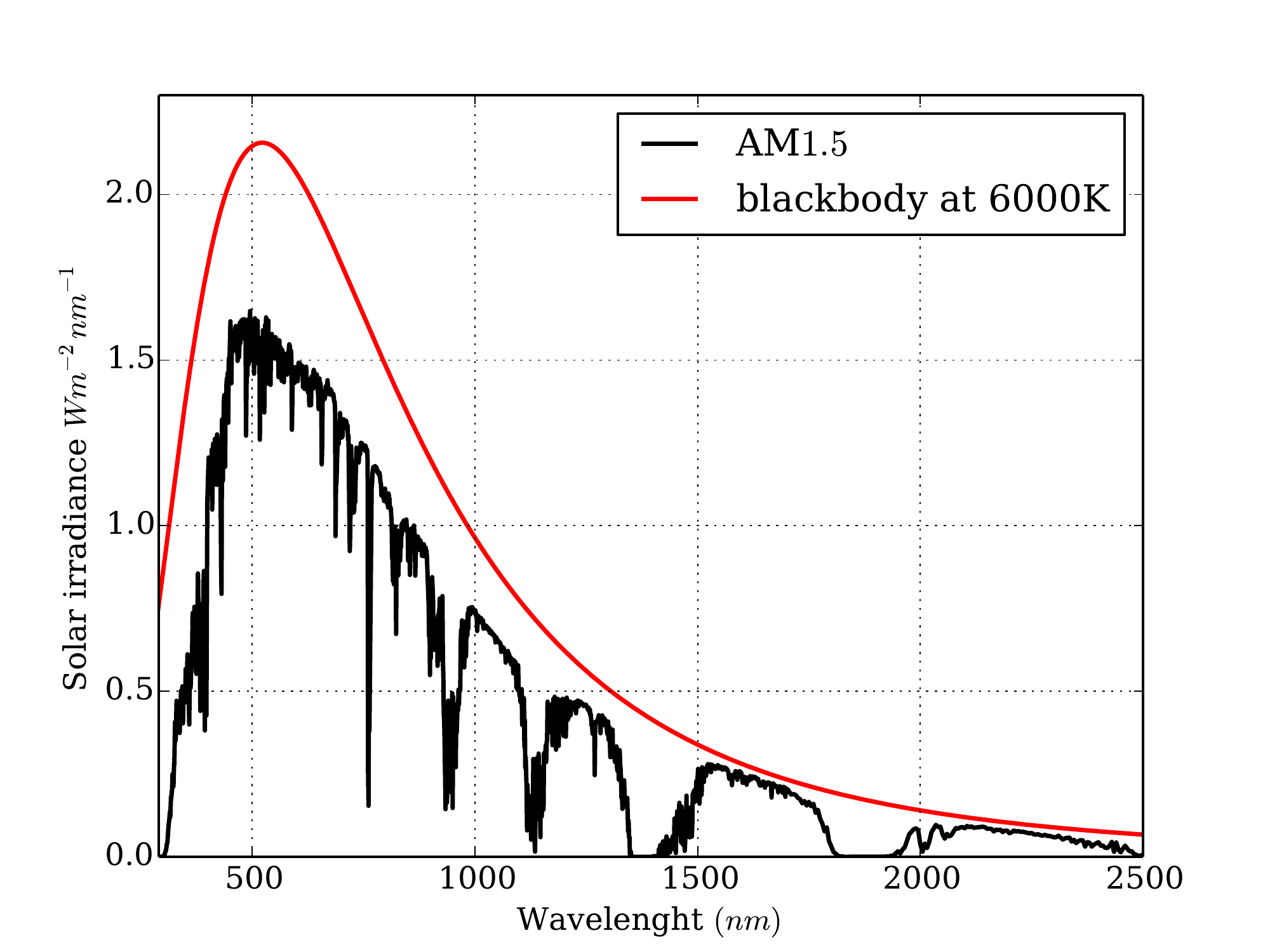}
\caption{Blackbody and AM1.5G solar spectral intensity.}\label{bbam}
\end{figure}

In this section, we used both the SQ  and SLME to estimate the solar cells efficiency of the 
materials. Figure~\ref{bbam} shows
two models of the solar spectral irradiance: the  blackbody model and the AM1.5G model  used as 
solar sources. The  AM1.5G spectrum  is not  as smooth as the blackbody prediction since it is an 
average distribution of the solar radiation depending on the day, the location and the path taken 
from the sun to the surface of the earth. 

\begin{figure}[hbtp]
\centering
\includegraphics[scale=0.42]{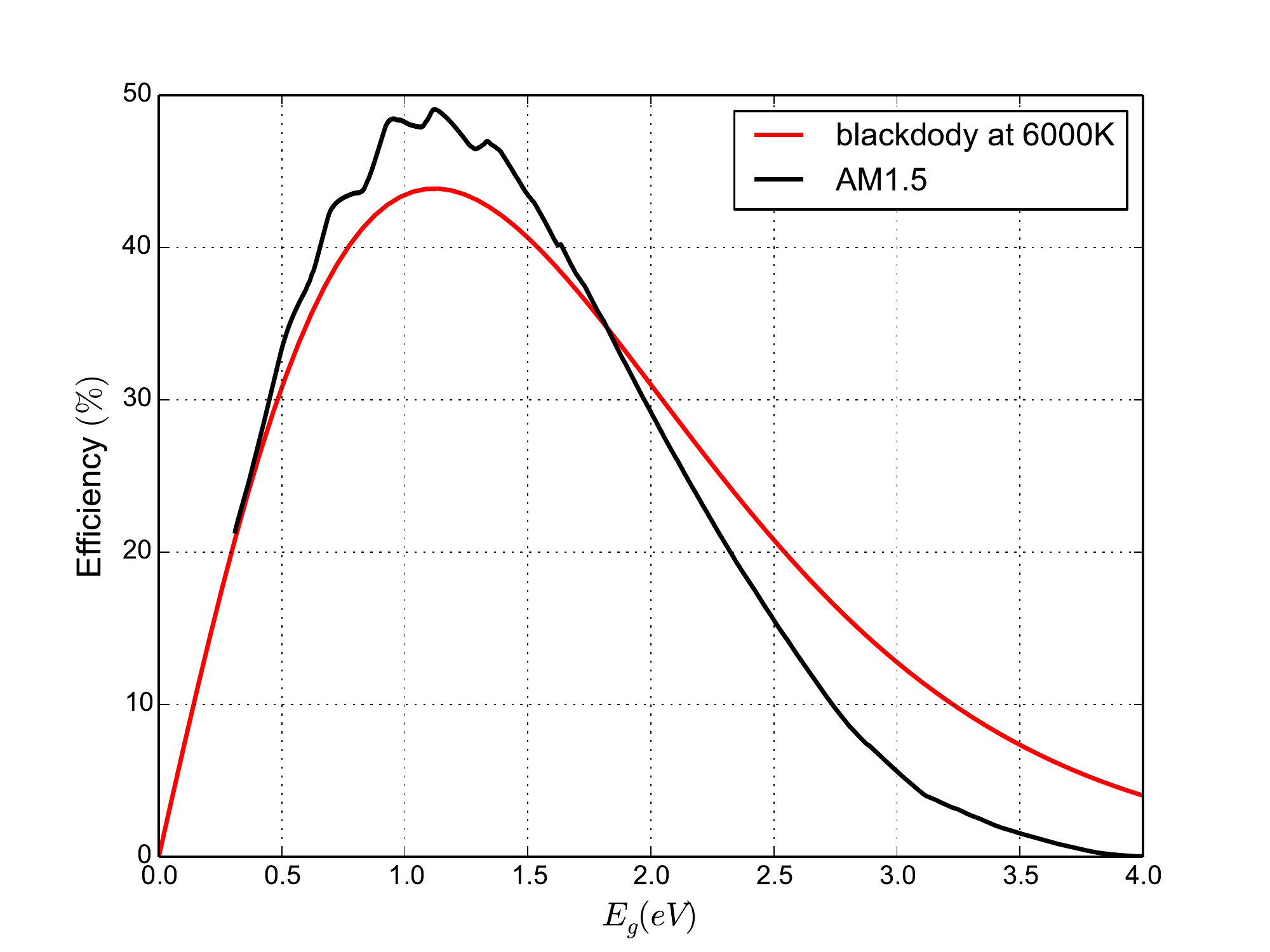}
\caption{Comparison between the blackbody at 6000 K and the AM1.5G efficiency 
limit.}\label{bb_sun-AM15}
\end{figure}

The Shockley and Queisser limit is a classic way of predicting the efficiency of the solar cells. 
Here,
the bandgap is the main parameter used to differentiate between the materials.   It can be observed 
from the SQ limit efficiency with the blackbody (BB)  spectra in Figure~\ref{bb_sun-AM15} that the 
maximum efficiency is about 43.87$\%$ for an energy of 1.12 eV.  It suggests that the best solar 
absorbers  should be those with bandgap in that range and this may justify why the existing 
technologies in photovoltaic are almost all based on silicon. We also notice that the maximum 
efficiency from the AM1.5G spectra is at 49.08$\%$. In addition, from 1.82 eV upwards, the number of 
photon absorbed by solar cell is smaller than that from the blackbody model.
Table~\ref{eff_am_bb} shows that AgAlS$_2$ and AgInS$_2$ have  respectively the least and the most efficient solar absorber materials from this study. 
For some materials, the difference in solar efficiency between the two solar models is important. For instance, there is a 66.6$\%$ difference between the BB and the AM1.5G solar irrandiance model in the case of AgAlS$_2$.
\begin{table}
[hbtp]
\centering%
\caption{solar efficiency($\%$) from SQ and SLME models using the BB and the AM1.5G spectrum}\label{eff_am_bb}
\resizebox{8 cm}{!}{
\begin{tabular}{cccccccccc}
\toprule
Materials& AgAlS$_2$&AgAlSe$_2$&AgAlTe$_2$&AgInS$_2$    \\ \toprule
SQ$@$BB &10.36&21.56&26.50&32.49 \\  \midrule
SQ$@$AM1.5G  & 3.46 & 16.26 & 23.08 & 31.65 \\	\midrule
 %recombination &\\	\midrule
SLME$@$AM1.5G  & 2.37 & 10.86 & 14.35 & 18.70 \\  \bottomrule
%recombination &\\ \bottomrule				           	 
\end{tabular}
}
\end{table}
An important factor to be considered while computing the efficiency of a solar cell absorber
is the recombination process. Recombination happens when an excited electron in the conduction band 
loses energy and falls into the valance band which is neutralized by the hole. 
Numerous recombination processes have been inventoried including radiative and non radiative 
recombination. Hence, all the generated electrons  are not always converted into current and the 
fraction of exact current is given by Equations~\ref{j} and~\ref{jprime}.
In addition to the recombination processes, another  factor which may influence the 
efficiency of a solar cell  material is the specific shape of the absorption
near the onset. In fact, having  strong absorption and a direct bandgap is not a
guarantee of a good solar cell material. Some materials with well positioned dipole forbidden  
direct transition lower than dipole allowed direct transitions might have a good 
efficiency~\cite{liping}. Based on the nature of the transition, there are four possible types of 
optical transitions namely: the direct allowed(da), the indirect allowed, direct forbidden and 
indirect forbidden transitions. 
The relation between the absorption ($\alpha$) and the incident photon can be used to
determine the nature of the optical transition following the Tauc's relation~\cite{Tauc1}:
\begin{align}
\alpha h \nu = \alpha_0\left(h\nu -E_g^{opt}\right)^n,
\end{align}  
where $E_g^{opt}$  is the optical gap  and $\alpha_0$ is the band tailing parameter.
Depending on the value of the power factor $n$, the transition can be  a direct allowed, an 
indirect allowed, a direct forbidden or an  indirect forbidden for $n=1/2, 2, 3/2$ or  $3$, 
respectively.   From the Tauc's plot fitting of the absorption as illustrated in Figure~\ref{f1}, we 
 found that all the materials studied have a direct allowed transition ($n=1/2$). These findings 
agree with the results of Liping et \textit{al.} \cite{liping}  except for AgAlTe$_2$ where they 
found  a direct-but-forbidden transition. The difference may probably be due to the fact that they 
determined the nature of the transition from the magnitude of matrix element square.
The recombination processes  together with  the nature of the optical transition are some of those 
factors that the classical SQ model does not take into consideration. Accounting for these two 
parameters as well as the absorptivity, the net current from Equation~\ref{j} becomes:
{\small
{\begin{align}\label{slmefinal}
&J=q\int_{E_g}^{\infty} \left[1-e^{-2\alpha(E)L}\right] AM1.5G(E) dE \nonumber\\
&-\dfrac{q\pi}{j_r}\left[ e^{(qV/k_BT_c)}-1\right] \int_{E_g}^{\infty}\left[1-e^{-2\alpha(E)L}\right]N_{ph}(E,T_c)dE,
\end{align}}%}
where $j_r=e^{-\Delta/k_BT}$ defines the fraction of the radiation electron-hole recombination with $\Delta=E_g^{da}-E_g$ (da= direct allowed).

In order to get the maximum power $P_{max}$ entering  in calculation of the solar cell efficiency $\eta$, Equation~\ref{slmefinal} should be integrated numerically throughout the AM1.5G solar spectral. 
The maximum power is obtained at the maximum power point tracking (MPPT) of a current-voltage (J-V) characteristic for a given value of the bandgap.% as illustrated in Figure~\ref{JV}

\begin{figure}[hbtp]
\centering
\includegraphics[scale=0.34]{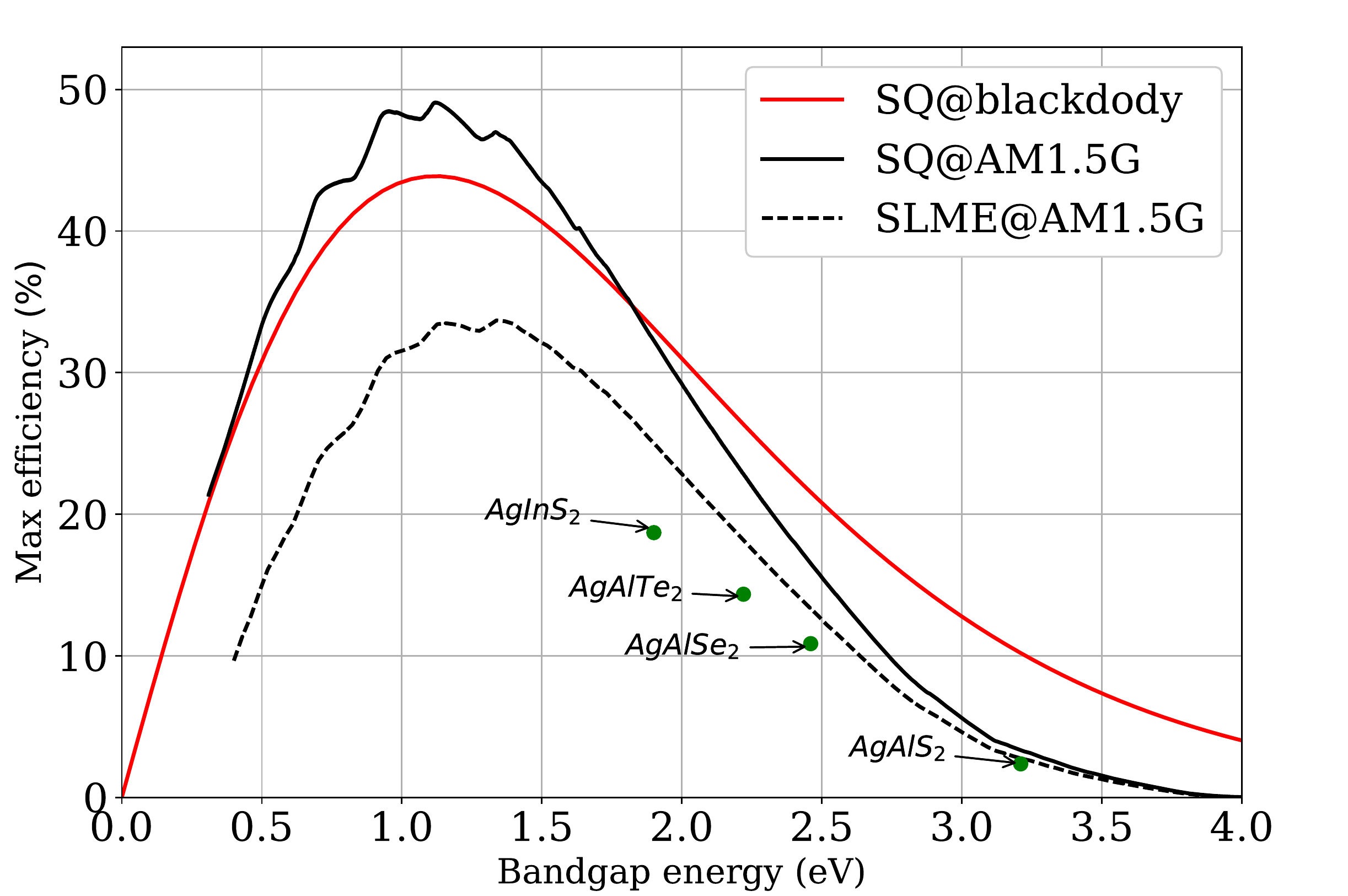}
\caption{Efficiency vs bandgap from different models.  The thickness of the thin film is set at  0.5$\mu m$.}\label{eff_gap}
\end{figure}
Repeating this procedure throughout the whole AM1.5G spectra leads to Figure~\ref{eff_gap} which depicts the 
dependence of efficiency with respect to the bandgap. The general trend is that the efficiency is 
lowered when accounting for the losses by recombination, the absorption spectra and the nature of 
the dipole transition. We obtained solar efficiencies of  2.37, 10.86, 14.35 and 18.70$\%$ for 
AgAlS$_2$, AgAlSe$_2$,  AgAlTe$_2$ and AgInS$_2$, respectively.  Recall that the absorption 
$\alpha(E)$ entering in  the calculation of absorptivity $a(E)$  was obtained at the BSE level of 
the approximation while the BSE calculations were built on top of the semi self-consistent GW 
approximation. There is not any previous work for the AgAlX$_2$ family, but a previous study on 
AgInS$_2$ predicted an efficiency about 20$\%$ ~\cite{liping,slme2}. There is a difference of about 
1$\%$ which could be attributed to different calculation methods. For instance, we used relaxed 
lattice parameters  whereas according to Ref.~\cite{liping}, the experimental lattices parameters 
were used.  It is known that the bandgap chalcopyrites  strongly depends on the internal structural 
parameters such as the anion  displacement  and the tetragonal distortion. Moreover, their 
absorption was obtained from HSE06 calculation with a scissor operator added to improve  the bandgap 
and the excitonic effect was not taken into consideration. Overall, the efficiency of the compounds 
of interest increases as the bandgap decreases for both the SQE and SLME. The low efficiency of 
AgAlS$_2$ relatively to others studied materials suggests that  it cannot be considered for single 
junction solar cell absorber.

\section{Conclusion}
We have reported the results of the  solar cell efficiency of AgMX$_2$ chalcopyrite materials based 
on the  optical and electronics properties from  first principles  calculations. Since G$_0$W$_0$ 
underestimates the bandgaps, we circumvented this problem by performing a semi self- consistent GW 
calculations for all the materials in this study. These results predicted that the bandgap of 
AgInS$_2$, AgAlS$_2$, AgAlSe$_2$ and AgAlTe$_2$ to be 1.9, 3.21, 2.46 and 2.22 eV, respectively. In 
order to accurately estimate the optical absorption in the materials studied, BSE   
equation in the Tamn-Anaconda approximations was used. We found that all the materials have a direct 
allowed dipole transition by means of the Tauc's plot fitting of the absorption spectra. The classic 
SQ and  SLME models were used for the calculation of the solar performance. We found that for a 
given absorber, the results of the  SQ  are relatively higher than those of  the SLME. This is due 
to the fact that the SQ model only  accounts for the bandgap neglecting the  recombination process, 
the absorption as well as the optical transition. We finally predicted  that the AgInS$_2$ with 
24.87$\%$ and AgAlS$_2$ with 2.37$\%$ have respectively the highest and the lowest solar cell 
efficiency.

\section*{Acknowledgements}
The Centre for High Performance Computing (CHPC), Cape Town-South Africa is Acknowledged for 
the computational resources. The authors are grateful to the  University of South Africa for 
financial support.

\bibliographystyle{ieeetr}
\bibliography{ref.bib}
\end{document}